\documentclass[amsmath,amssymb,groupedaddress,superscriptaddress,lengthcheck,aps,prl] {revtex4-1}
\usepackage{gensymb}
\usepackage{graphics}
\usepackage{graphicx}   
\usepackage{bm}
\usepackage[latin1]{inputenc}
\usepackage{textcomp}
\usepackage{float}

\begin{document}

\title{Optical emission spectroscopy study of competing phases of electrons
    in the second Landau level}
\author{A. L. Levy}
	\email{all2143@columbia.edu }
	\affiliation{Department of Physics, Columbia University, New York, NY 10027, USA}
\author{U. Wurstbauer}
	\affiliation{Walter Schottky Institut and Physik-Department, Technische Universit{\"a}t M{\"u}nchen, Am Coulombwall 4a, 85748 Garching, Germany}
        \affiliation{Nanosystems Initiative Munich (NIM), Munich, Germany}
\author{Y. Y. Kuznetsova}
	\affiliation{Department of Physics, Columbia University, New York, NY 10027, USA}
\author{A. Pinczuk}
	\affiliation{Department of Physics, Columbia University, New York, NY 10027, USA}
	\affiliation{Department of Applied Physics and Applied Mathematics, Columbia University, New York, NY 10027, USA}
\author{L.~N.~Pfeiffer}
	\affiliation{Department of Electrical Engineering, Princeton University, Princeton, NJ 08544, USA}
\author{K.~W.~West}
	\affiliation{Department of Electrical Engineering, Princeton University, Princeton, NJ 08544, USA}
\author{M. J. Manfra}
	 \affiliation{Department of Physics and Astronomy, Birck Nanotechnology Center, Purdue University, West Lafayette, IN 47907, USA}	
	 \affiliation{School of Materials Engineering, Birck Nanotechnology Center, Purdue University, West Lafayette, IN 47907, USA}	
	 \affiliation{School of Electrical and Computer Engineering, Birck Nanotechnology Center, Purdue University, West Lafayette, IN 47907, USA}	
\author{G. C. Gardner}
	 \affiliation{School of Materials Engineering, Birck Nanotechnology Center, Purdue University, West Lafayette, IN 47907, USA}
\author{J. D. Watson}
	 \affiliation{Department of Physics and Astronomy, Birck Nanotechnology Center, Purdue University, West Lafayette, IN 47907, USA}

	\date{\today}
	
\begin{abstract}
Quantum phases of electrons in the filling factor range $2 \leq\nu\leq 3$ are probed by the weak optical emission from the partially populated second Landau level and spin wave measurements. Observations of optical emission include a multiplet of sharp peaks that exhibit a strong filling factor dependence. Spin wave measurements by resonant inelastic light scattering probe breaking of spin rotational invariance and are used to link this optical emission with collective phases of electrons. A remarkably rapid interplay between emission peak intensities manifests phase competition in the second Landau level. 

\end{abstract}


\maketitle

Ultra-clean two dimensional electron systems in the presence of high perpendicular magnetic fields $B$ are a source of unexpected and fascinating quantum many-body physics that arises from  the strong electron interactions combined with a reduction in dimensionality. When $B$ is high enough for all electrons to occupy the lowest (N=0) Landau level (LL), the many-electron system forms liquids of the fractional quantum Hall effect (FQHE). When $B$ is such that electrons fill states in higher ($\textnormal{N}\geq2$) LL's, electrons form quantum phases referred to as stripe and bubble phases, which lead to transport anisotropy and reentrant integer quantum Hall effect (RIQHE) states \cite{Lilly1999,Du1999,Pan1999}. 
The unique electron-electron interactions in the N=1 LL  result in the presence of RIQHE states and stripe phases in addition to even- and odd-denominator FQHE states  \cite{Pan1999,Csathy2005}. FQHE states in the second (N=1) LL exhibit even-denominator states such as the one at $\nu=5/2$~ \cite{Willett1987,Eisenstein1988}, which is predicted to have non-Abelian excitations~\cite{Moor1991,Reza2000,Lev2007,Lee2007,Pete2008,Stor2011,Stern2010,Reza2011},  has recently been studied by NMR \cite{Tiem2012, MStern2012}, by light scattering methods~\cite{Rhone2011,Wurstbauer2013}, and in two-subband systems~\cite{Nuebler2012}. It has been predicted that the less studied FQHE state at $\nu=2+1/3=7/3$ could possess exotic quasiparticles in which composite fermions are dressed by a cloud of neutral excitations \cite{balram2013}. Since FQHE liquids as well as bubble and stripe phases can serve as ground states, the N=1 LL is home to a striking competition between quantum phases~\cite{Deng2012}.

The interplay of anisotropic phases with FQHE liquids in the second LL has been studied by introduction of in-plane magnetic fields \cite{Pan1999,Xia2010,Xia2011,Shay2013,Friess2014}.  
These experiments provide evidence that anisotropic smectic- or nematic-like phases with broken full rotational invariance coexist with quantum Hall liquids \cite{Shkl1996,Musa1996,Moess1996,Fert1999,Frad2000}. The large anisotropy induced in the system at the FQHE states at $\nu=5/2$ and $\nu=7/3$ by relatively small in-plane magnetic fields \cite{Xia2010,Xia2011,Shay2013} supports interpretations in terms of a new state of electron matter with FQHE states that occur in the  environment of a nematic stripe phase \cite{Frad2000,Mull2011,Qiu2012}.  

\begin{figure}[htpb]
\centering
\includegraphics[width=3.4in]{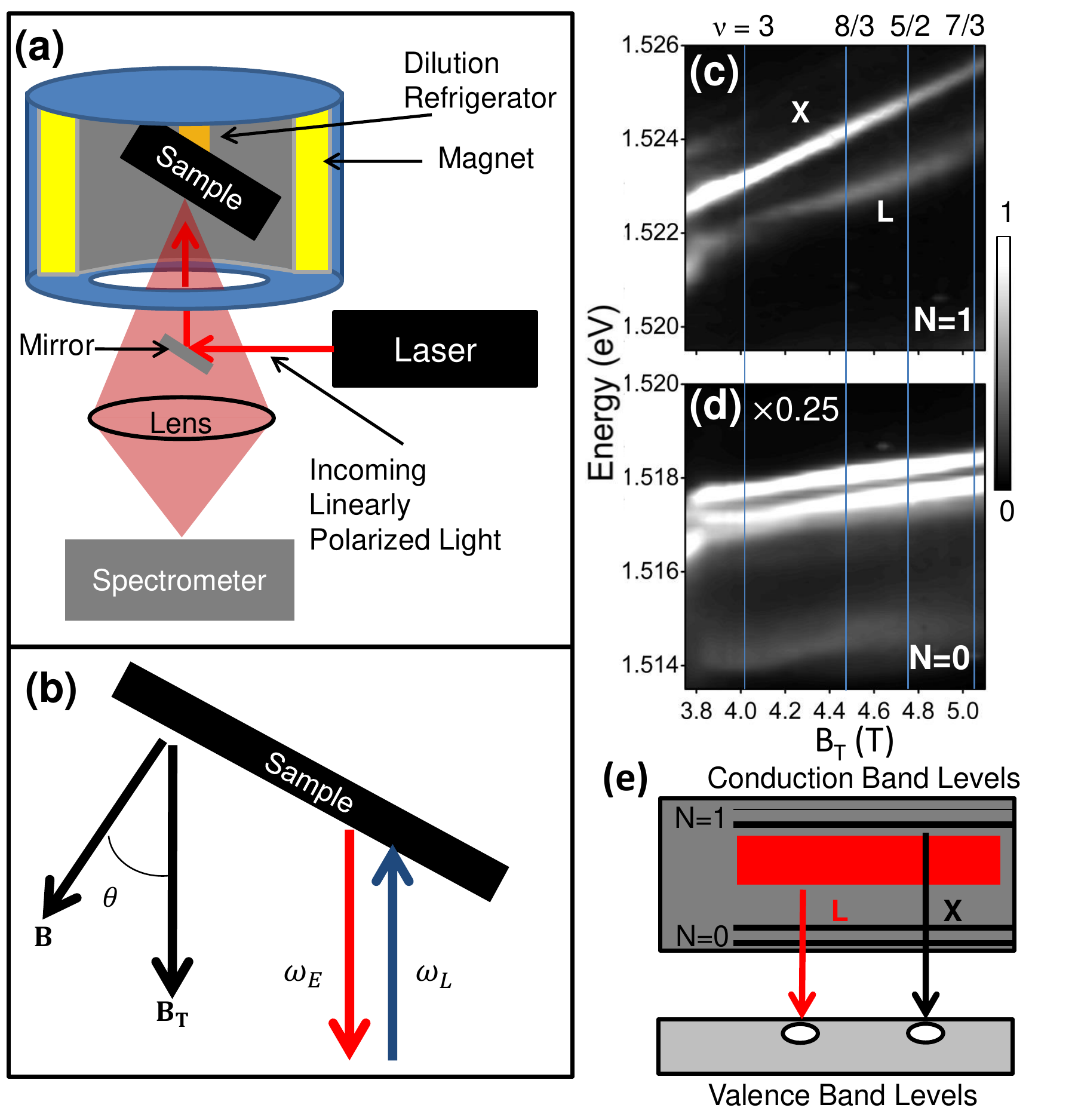}
\caption{(a) Experimental set-up with bottom optical access in the dilution refrigerator. (b) Schematic description of  the experimental geometry showing incident and emitted photons and the tilt angle $\theta$ of the sample. The total magnetic field $B_{T}$ and the perpendicular component $B$ are also shown. (c) Energy vs. $B_{T}$  observed in optical emission spectra from the  N=1 LL in the filling factor range $2\leq\nu\leq3$ for sample A. The intensity is shown in grayscale. The band labeled X is linearly dispersed in $B_{T}$. L is the red-shifted optical emission that is considered in the main text. (d)  Energy vs. $B_{T}$ plot for optical emission spectra from the  N=0 LL in the range $2\leq\nu\leq3$ for sample A.  (e) Schematic description of optical emission transitions that originate in the N=1 LL.} 
\label{fig:fig1}
\end{figure}

We report optical emission experiments that probe quantum phases that emerge in the second LL of an ultra-clean 2D electron system. The optical recombination is from transitions across conduction to valence band states from electrons that partially populate the N=1 LL. This emission, while much weaker than the one originating in the N=0 LL (see Figs.~\ref{fig:fig1}(c) and (d)), displays a marked dependence on filling factor, which uncovers competing and overlapping quantum phases in the range $2<\nu<3$.

Links between optical emission and emerging quantum phases are established by comparing optical emission with the long-wavelength spin wave obtained by resonant inelastic light scattering (RILS).
At ferromagnetic quantum Hall states such as $\nu = 3$, all spins are aligned, and the long wavelength spin wave occurs at the bare Zeeman energy, in agreement with Larmor theorem \cite{Kallin1984}. The departure from the Larmor theorem for $\nu < 3$ is regarded as the evidence of formation of spin textures that break the full rotational invariance of the 2D electron system due to the combined effects of Coulomb interactions and disorder \cite{Drozdov2010,Rhone2011,Wurstbauer2013}. 

The emission from the N=1 LL displays two major components, a singlet with linear magnetic field dependence and a red-shifted multiplet with a striking dependence on filling factor. An investigation of links between optical emission and spin waves in RILS spectra allows to link the peaks in the red-shifted optical emission to quantum phases in the N=1 LL.  The wide ranges of filling factors over which these phases exist together with the absence of a clear temperature dependence for T $\leq$ 300~mK indicates that these are not FQHE or RIQHE phases, but, we surmise, phases that coexist with them.

The filling factor dependence of the red-shifted optical emission is particularly striking in the filling factor range $2 \lesssim \nu \lesssim 2.5$, where three distinct peaks display rapid changes in intensity with magnetic field in a narrow filling factor range. 
Softening of the spin wave from the Zeeman energy in this filling factor range is similar to the effect reported in the filling factor range $2/3<\nu<1$, which was interpreted as arising from the appearance of spin textures in the ground state \cite{Gallais2008}. Measurements of the spin wave by RILS thus allow to probe the spin rotational invariance of the competing phases observed through optical emission.

The 2D electron system is realized in two samples each with a symmetrically doped single GaAs/AlGaAs quantum well of width 300~\AA~\cite{Pfeiffer2003,Manfra2014}.  The charge carrier density in the lower density sample A is $2.92~\times~10^{11}$~cm$^{-2}$, measured in transport experiments, and the carrier mobility is $23.9~\times~10^{6}$~$\textnormal{cm}^2/\textnormal{Vs}$ (at 300~mK). The higher density sample B has a density of $3.2\times10^{11}~\textnormal{cm}^{-2}$ and mobility of $20\times10^{6}$~cm$^2/$Vs (at 300 mK). Samples are mounted on the cold finger of a $^3$He$/^4$He dilution refrigerator operating at a base temperature below 40~mK and placed in the bore of a 16~T superconducting magnet. Bottom windows are employed for spectroscopy (Fig.~\ref{fig:fig1}(a)). The optical emission spectra are excited by a tunable Ti:Sapphire laser at an incident power below $10^{-4} \rm W/\rm cm^{2}$ and recorded in the backscattering geometry  shown in Fig.~\ref{fig:fig1}(b). Laser heating at this power density keeps the electron gas temperature below 100~mK at the base temperature of the dilution refrigerator, as demonstrated in Ref.~\cite{Kang2000}. The excitation wavelength of 800~nm is at a photon energy close to the fundamental optical gap of the GaAs quantum well. The sample is tilted at an angle $\theta = 20\degree$. The resulting small in-plane component of the magnetic field allows for well-defined FQHE states at $\nu~=~5/2$ and $\nu~=~7/3$ and also anisotropic phases in the second LL~\cite{Pan1999,Csathy2005,Xia2010,Xia2011}. The filling factor is identified by the strong spin wave in the polarized $\nu=3$ state as described in the supplementary online information \cite{suppl}.

\begin{figure}[htpb]
\centering
\includegraphics[width=3.4in]{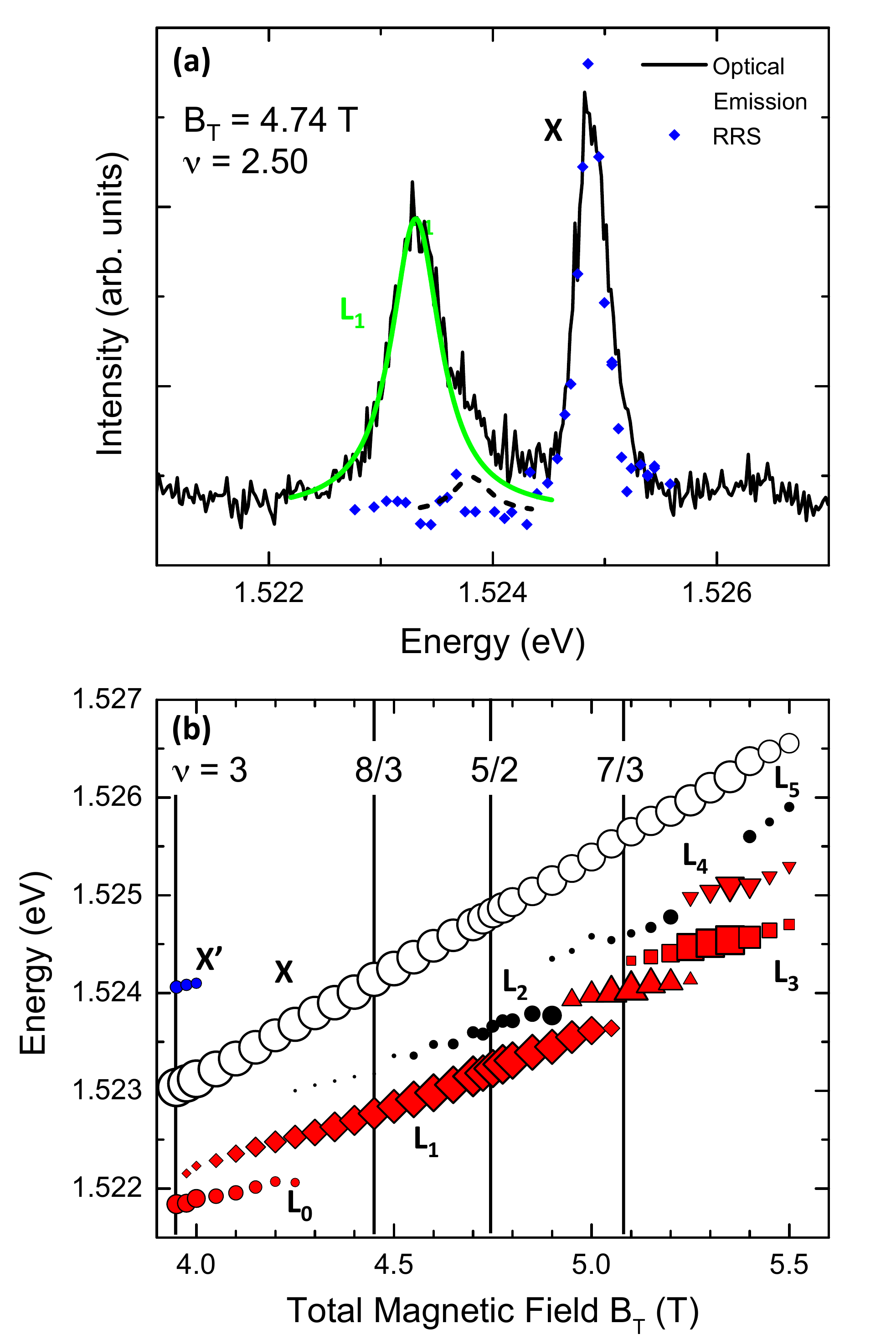}
\caption{(a) RRS results overlapped with optical emission for $\nu=2.50$ from sample A. (b) Energy of the bands in the optical emission from sample A from the N=1 LL as a function of total magnetic field $B_T$. The area of each data point is proportional to the integrated intensity found from a Lorentzian fit (except in the case of L$_0$ and L$_4$, which appear Gaussian) such as the green curve in (a) and normalized by the electron population of the N=1 LL. The black filled circles indicate low intensity emission with higher uncertainty on its energy, such as the black dashed curve in (a).}
\label{fig:fig2}
\end{figure}

The optical emission is well represented by multiple Lorentzians with varying amplitude and nearly constant width (the width itself depending on the particular peak). The results of such line shape analysis in the range $2\leq\nu\leq3$ are summarized in Figure~\ref{fig:fig2}(b), which presents peak energies as a function of total magnetic field $B_T$. The area of each data point is proportional to the integrated intensity of the peak found from a Lorentzian fit such as shown in Fig.~\ref{fig:fig2}(a) and normalized by the electron population of the N=1 LL. 

Figures~\ref{fig:fig1}(c) and (d) summarize optical emission results in the range $2\leq\nu\leq3$. The emission doublet from the N=0 LL (Fig.~\ref{fig:fig1}(d)) is similar to those reported in previous studies~\cite{Gravier1998,MStern2010}. The focus here is on the much weaker optical recombination due to transitions that originate from partially populated states in the N=1 LL shown in Fig.~\ref{fig:fig1}(c), which displays two major features labelled as X and L. 
This result is markedly  different from the N=0 emission spectra in a range $\nu\leq1$, where bands disperse linearly in $B$ and display oscillation in energy as a function of $\nu$~\cite{Byszewski2006,Nomura2014}.

Figure~\ref{fig:fig2}(a) presents a typical optical emission spectrum and resonant Rayleigh scattering (RRS) at $\nu=2.50$. In the partially populated N=1 LL, RRS identifies the energy of the excitonic transitions between the partially populated conduction band and the valence band \cite{Rhone2011}. 
The RRS measurements reveal the X peak as resulting from excitonic transitions.
The energy of the singlet X band has a linear dependence on the perpendicular component of the magnetic field $B$ with a slope of $2.39~\pm~0.05$~meV$/$T, illustrated in Fig.~\ref{fig:fig2}(b). This value of the slope is close to that of free electrons in GaAs in the N=1 LL. Such magnetic field dependence indicates that the X emission arises from optical transitions at energies that are modified from single particle transition energies of conduction and valence LLs by excitonic interactions and weak coupling to the electron system. 
The red-shifted L emission is a multiplet structure (Fig.~\ref{fig:fig2}(a)) that exhibits a strong dependence on filling factor (Fig.~\ref{fig:fig2}(b)). The optical transitions for the L peaks are shown in Fig.~\ref{fig:fig1}(e) as red-shifted from single-particle conduction states.  The RRS measurements in Fig.~\ref{fig:fig2}(a) show that the absorption edge is at the X peak, suggesting that the recombination responsible for the L multiplet consists of lower energy electron states than the X peak.

\begin{figure}[htpb]
\centering
\includegraphics[width=3.4in]{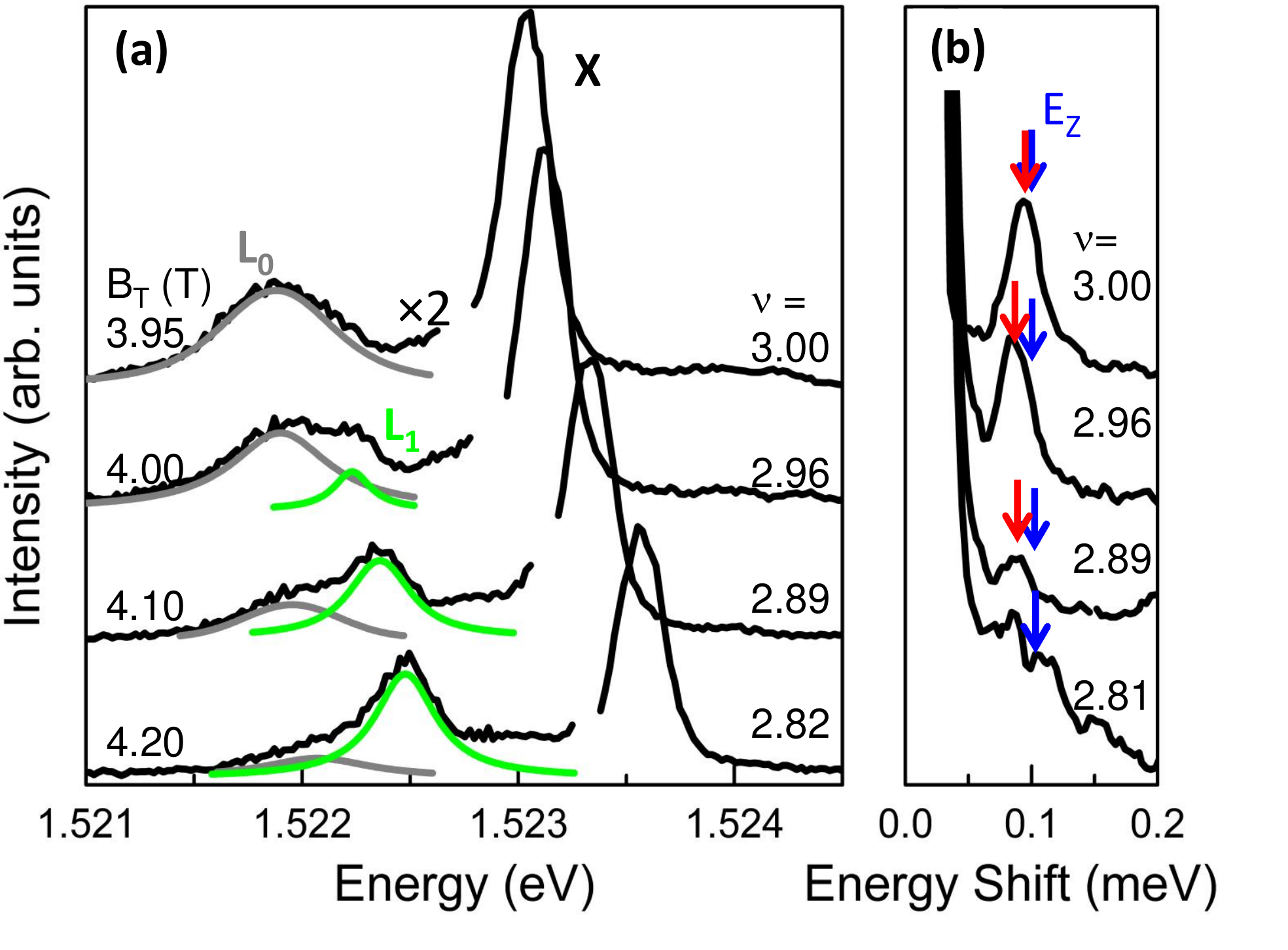}
\caption{(a) Optical emission and (b) RILS spectra from sample A for filling factors close to $3$. The color curves in (a) are fits with Lorentzian functions. The observed spin wave in (b) is indicated with a red arrow and compared to the Zeeman energy (blue arrow). Data shown in (b) was collected during a different cooldown of the dilution refrigerator~\cite{Wurstbauer2013}, which results in a small difference in magnetic fields that achieve the same filling factors as (a).}
\label{fig:fig3}
\end{figure}

Figure~\ref{fig:fig3} establishes the link between the red-shifted L emission peaks and electron phases near $\nu = 3$. The interplay between the L peaks (Fig.~\ref{fig:fig3}(a)) correlates with a softening and collapse of the Zeeman mode (Fig.~\ref{fig:fig3}(b)).  At  $\nu=3$, the L emission consists of a singlet peak labeled L$_{0}$ (Fig.~\ref{fig:fig3}(a)). The rapid reduction of the L$_{0}$ intensity  with decreasing filling factor and simultaneous softening of the spin wave clearly  indicates that the L$_{0}$ emission is characteristic of the integer QHE state at $\nu=3$. 
Figures~\ref{fig:fig2}(b) and~\ref{fig:fig3}(a) illustrate the emergence of a new peak L$_{1}$ around $\nu=2.96$, which becomes the dominant feature of the L emission for $\nu\lesssim2.9$. 
Figure~\ref{fig:fig3}(b) shows a strong Zeeman mode at $\nu = 3$ that rapidly decreases in energy and collapses as the L$_1$ peak gains intensity. 
The correlation between the emission and spin wave spectra links the appearance of the L$_{1}$ band to the emergence of a new phase in the partially populated N=1 LL.
The softening and collapse of the spin wave away from $\nu = 3$ indicates the presence of spin textures that break the full rotational invariance necessary to support spin waves at the Zeeman energy. 

\begin{figure}[htpb]
\centering
\includegraphics[width=3.4in]{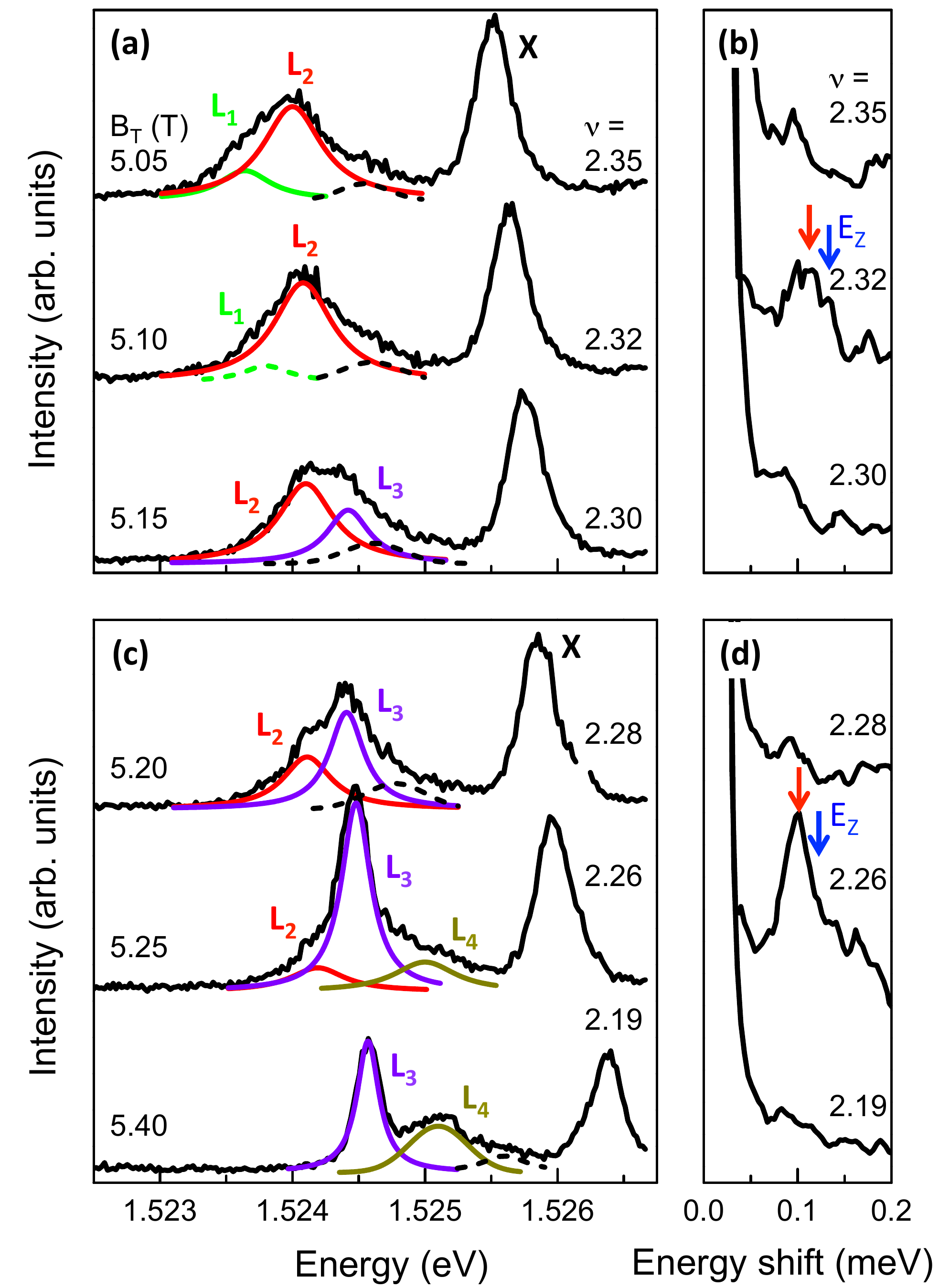}
\caption{(a) Optical emission and (b) RILS spectra from sample A for filling factors close to $7/3$. (c) Optical emission and (d) RILS spectra from sample A for a filling factor range where peak L$_3$ is dominant. The color curves in (a,c) are fits with Lorentzian functions. The observed spin wave in (b,d) is indicated with a red arrow and compared to the Zeeman energy (blue arrow).}
\label{fig:fig4}
\end{figure}

The most striking feature of the L multiplet is the interplay between the intensities of L$_{1}$, L$_{2}$ and L$_{3}$ peaks in the vicinity of $\nu=7/3$ (Fig.~\ref{fig:fig4}). 
As $B_T$ increases and $\nu$ approaches $7/3$, the L$_{1}$ component loses intensity and disappears from the spectra for $\nu\lesssim2.32$. Simultaneously, the  L$_{2}$ band, which becomes well-defined for $\nu<5/2$ (Fig.~\ref{fig:fig2}(b)), increases in intensity, as seen in Fig.~\ref{fig:fig4}(a). A similar competition is seen in the results presented in Fig.~\ref{fig:fig4}(c), where the intensity of  L$_{3}$ increases sharply as the intensity of L$_{2}$ quickly collapses. 
RILS spectra display a recovery of the long-wavelength spin waves near $\nu=7/3$, where the intensity of L$_{2}$ is the highest, and at $\nu=2.26$, where L$_{3}$ dominates the multiplet (Fig.~\ref{fig:fig4}(b,d) and \cite{Wurstbauer2013}). The discernible softening of the spin wave from the Zeeman energy near $\nu=7/3$ (Fig.~\ref{fig:fig4}(b)) and $\nu=2.26$ (Fig.~\ref{fig:fig4}(d)) is similar to the one observed at $\nu<3$ (Fig.~\ref{fig:fig3}(b)). A similar interpretation to explain the results in Fig.~\ref{fig:fig4} suggests that the softening of the spin wave is evidence that the phases responsible for the  L$_2$ and L$_3$ emission bands, similar to L${_1}$, possess spin textures that break the full rotational invariance. This interpretation is consistent with  results from anisotropic transport at $\nu=7/3$ \cite{Xia2011}. 
The L$_2$ emission fully dominates the red-shifted multiplet near $\nu=7/3$ (Fig.~\ref{fig:fig4}(a)) and is thus associated with a quantum phase that is dominant near $\nu=7/3$.

We vary the temperature to gain additional insights into the nature of the observed quantum phases. The temperature dependence of the optical emission appears to be negligible for T $<$ 300 mK for the entire range $2<\nu<3$ (see Fig. S4 in the supplementary information \cite{suppl}, where sample B is studied). For a large range of filling factors, optical emission does not exhibit a discernible temperature dependence below 650~mK, whereas there is a clear temperature dependence at certain filling factors, notably $\nu = 2.32$ (Fig. S4(b)), for 300 mK $<$ T $<$ 650 mK. The fits suggest a competition between L$_{2}$ and L$_{3}$, with the L$_{3}$ gaining intensity and L$_{2}$ shrinking with increasing temperature.  This temperature dependence is significantly weaker than that of FQHE and RIQHE \cite{Kumar2010} and is more similar to the temperature dependence of anisotropic transport at $\nu=7/3$ \cite{Xia2011}. 

The exploration of optical emission from the partially populated N=1 LL offers new insights into exotic quantum phases that emerge in the filling factor range $2\leq\nu\leq3$. 
The anomalous spin waves that correlate with the presence of L$_1$, L$_2$, and L$_3$ emission bands break Larmor theorem, indicating spin textures that lack  full spin rotational invariance. 
These results support a conceptual framework in which the bands of the L-multiplet are associated with distinct phases in the partially populated N=1 LL and the interplay in the peak intensities demonstrated in Figs.~\ref{fig:fig4} (a,c) is understood as revealing a sharp competition between phases that occurs near filling factors $7/3$ and 2.26.
The rapid changes that occur in the L-multiplet for filling factors near the FQHE state at $\nu=7/3$ suggest a striking competition between quantum ground states that are tuned by remarkably small changes in filling factor. The results demonstrate that optical methods form a powerful tool for the identification and study of exotic quantum phases of electrons in the partially populated N=1 LL.

The work at Columbia is supported by the National Science Foundation Division of Materials Research under  Awards DMR-1306976 and DMR-0803445, and by the Alexander von Humboldt Foundation. The research at TUM was supported by the Nanosystems Initiative Munich (NIM). The molecular beam epitaxy growth and transport characterization at Princeton University was supported by the Gordon and Betty Moore Foundation under Award GMBF-2719 and by the National Science Foundation, Division of Materials Research, under Award DMR-0819860. The molecular beam epitaxy growth and transport measurements at Purdue are supported by the U.S. Department of Energy, Office of Basic Energy Sciences, Division of Materials Sciences and Engineering under Award DE-SC0006671. We thank Dov Fields for analysis of data, Sheng Wang for insightful discussions, and A. F. Rigosi for support during measurements.

\nocite{Pinczuk1993}
\nocite{Goldberg1988}
\nocite{Hirjibehedin2003}
\nocite{Kumar2010}

\onecolumngrid
\newpage
\appendix

\section{SUPPLEMENTARY INFORMATION}

\section{Identification of the filling factor}

\begin{figure}[H]
\centering
\includegraphics[width=5.9in]{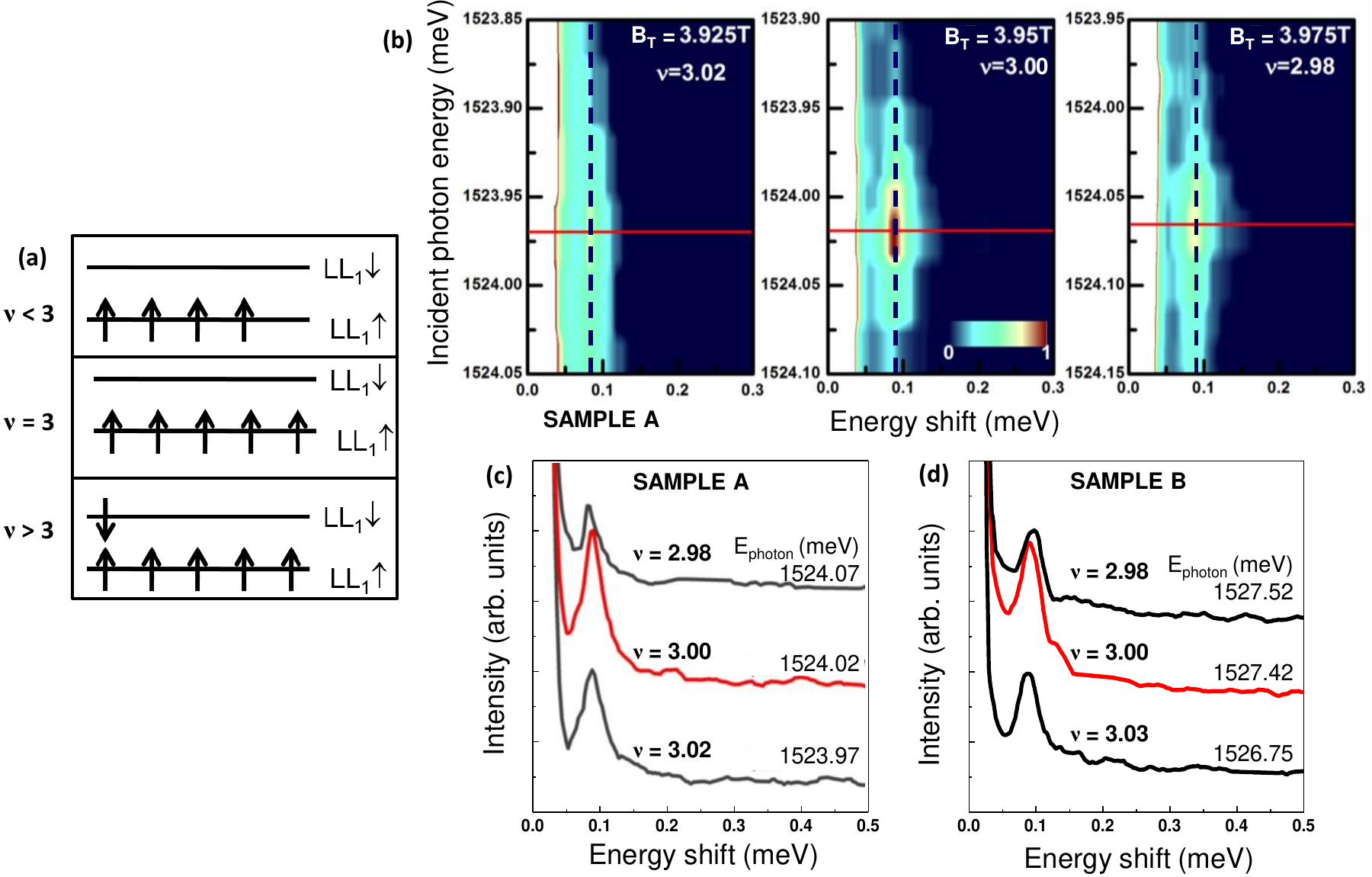}
\caption{(a) Configuration of spins around $\nu = 3$. (b) Intensities of RILS from sample A as a function of incident photon energy at filling factor $\nu = 3.02$, $3.00,$ and $2.98$. (c) Filling factor dependence of the resonantly enhanced SW mode around $\nu = 3$ for sample A. The SW intensity is significantly reduced for small changes in filling factor away from $\nu = 3$. (d) Intensities of RILS from sample B as a function of incident photon energy at filling factor $\nu = 3.03$, $3.00$, and $2.98$.}
\label{figS1}
\end{figure}

The magnetic field for filling factor $\nu = 3$ can be determined directly by RILS measurements from states in the N=1 LL.
For the fully spin polarized state at $\nu = 3$, the lowest-lying collective excitation mode is a spin reversal mode, a so called spin wave (SW) mode. As shown in Figure~\ref{figS1}(a), the spin wave gets weaker in both directions away from $\nu = 3$ due to the lower number of electrons in the N=1 LL for $\nu < 3$ and to a reduction of spin polarization for $\nu > 3$. 

RILS measurements from sample A at filling factors $\nu = 3.02$, $3.00$, and $2.98$, displayed in Fig.~\ref{figS1}(b), exhibit strong resonance. Resonant enhancement of spin wave mode at $E_Z$ is achieved only under extreme resonant conditions in a very narrow range of incident photon energies. In Fig.~\ref{figS1}(c), the most resonantly enhanced RILS spectra are compared for filling factors in close vicinity to $\nu = 3$. Minor changes of the filling factor significantly lower the intensity of the resonantly enhanced SW mode as shown in Fig.~\ref{figS1}(c). 
Fig.~\ref{figS1}(d) shows analogous resonantly enhanced RILS spectra for sample B. 
Remarkably, even for the very robust integer quantum Hall state at $\nu = 3$, there is a significant filling factor dependence of the mode intensity in a much narrower filling factor range than can be expected from the width of the quantum Hall plateau in magneto-transport measurements under similar conditions.

The precise determination of the magnetic field for $\nu = 3$ from the maximum of the SW mode intensity in RILS measurements allows us to calculate the magnetic fields for any filling factor.


\section{Optical emission studies of sample B}

\begin{figure}[H]
\begin{minipage}{.5\textwidth}
\centering
\includegraphics[width=3.1in]{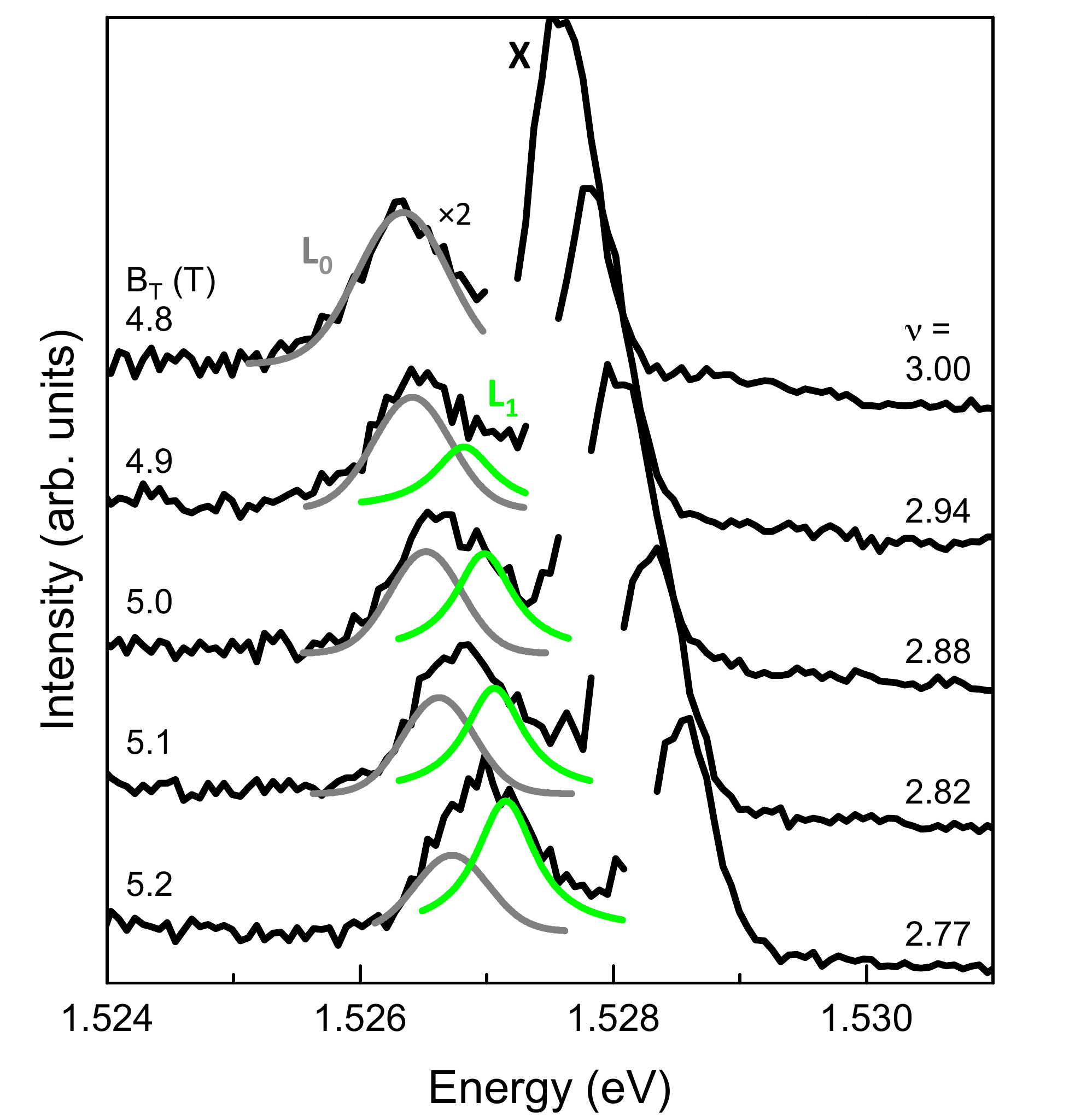}
\caption{Optical emission from sample B near $\nu=3$. $T$ = 42 mK.}
\label{figS2}
\end{minipage}
\begin{minipage}{.5\textwidth}
\includegraphics[width=3.1in]{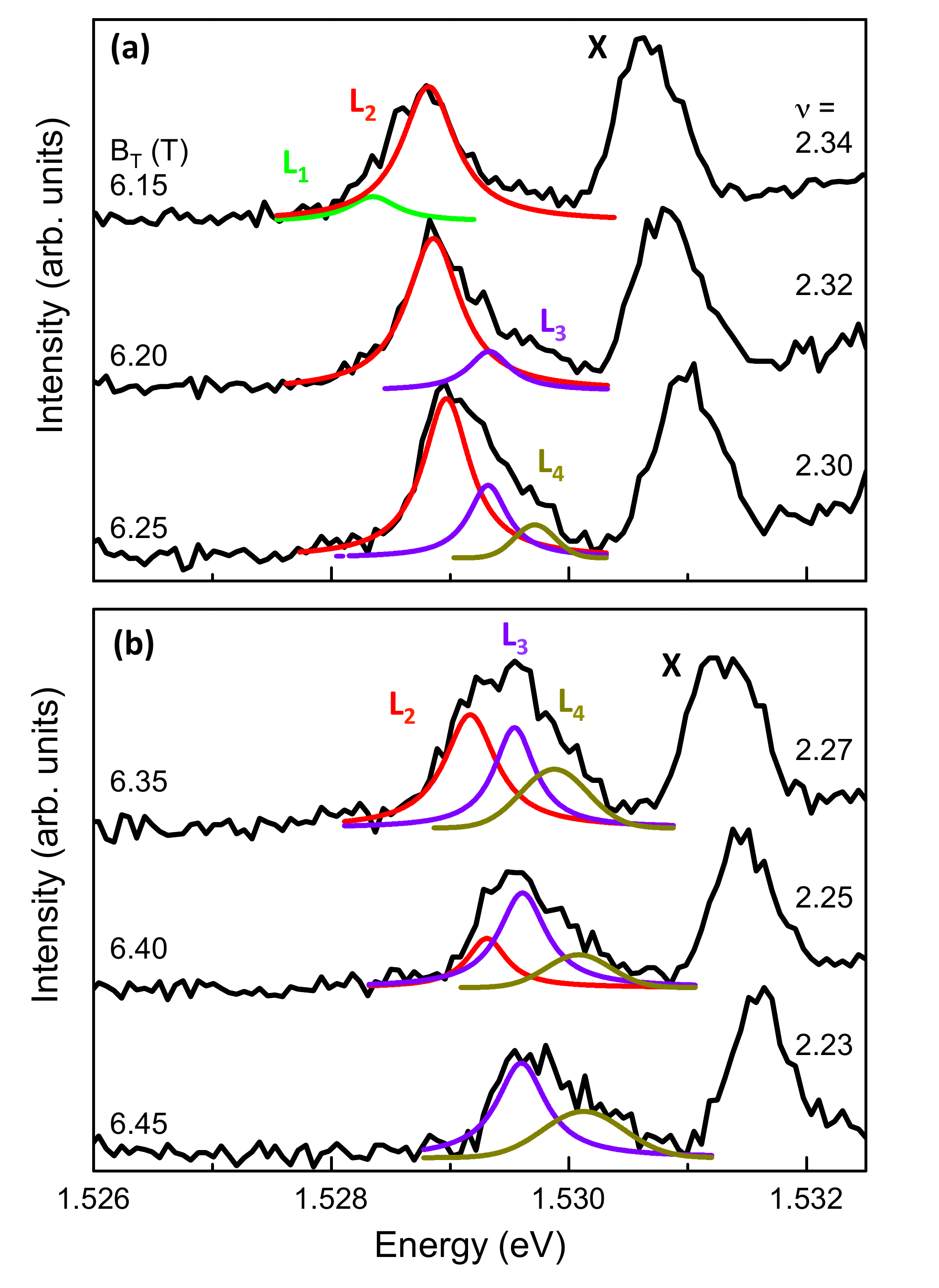}
\caption{Optical emission from sample B near $\nu=7/3$ (a) and lower (b). $T$ = 42 mK.}
\label{figS3}
\end{minipage}
\end{figure}

Optical emission measurements of sample B reveal an X peak whose energy disperses linearly with magnetic field and an L multiplet that exhibits non-trivial filling factor dependence. Figure~\ref{figS2} shows the emission from sample B near $\nu=3$. While the fits of the emission spectra of Sample B are less precise than those in sample A, the L-multiplet emission fro m sample B is clearly asymmetric and changes shape and width considerably with filling factor, allowing us to fit it with multiple peaks of nearly constant width. The decay of L$_{0}$ and simultaneous increase of L$_{1}$ for $\nu<3$ are consistent with the observations for sample A.  

Figure~\ref{figS3} shows the optical emission spectra from sample B at filling factors near to $7/3$. This filling factor dependent behavior of the peaks in the L multiplet resembles that of Sample A. In Sample B, similarly to Sample A,  
L$_3$ emerges at $\nu\approx$2.32 and becomes the most the dominant peak at $\nu\approx$ 2.27. Other parallels between the spectra from both samples include the fact that L$_{3}$ is the narrowest peak in the L multiplet, the filling factors at which peaks L$_2$ and L$_4$ are most prominent, the filling factors at which they disappear, and the L-peaks' shapes and widths relative to one another.
The similarities between the emission spectra from samples A and B point to origins of the L peaks that are fundamental to physics of the N=1 LL and not idiosyncratic to either sample, as these samples had different electron densities and were grown in different molecular beam epitaxy chambers.


\section{Temperature dependence of optical emission}

\begin{figure}[H]
\centering
\includegraphics[width=5.5in]{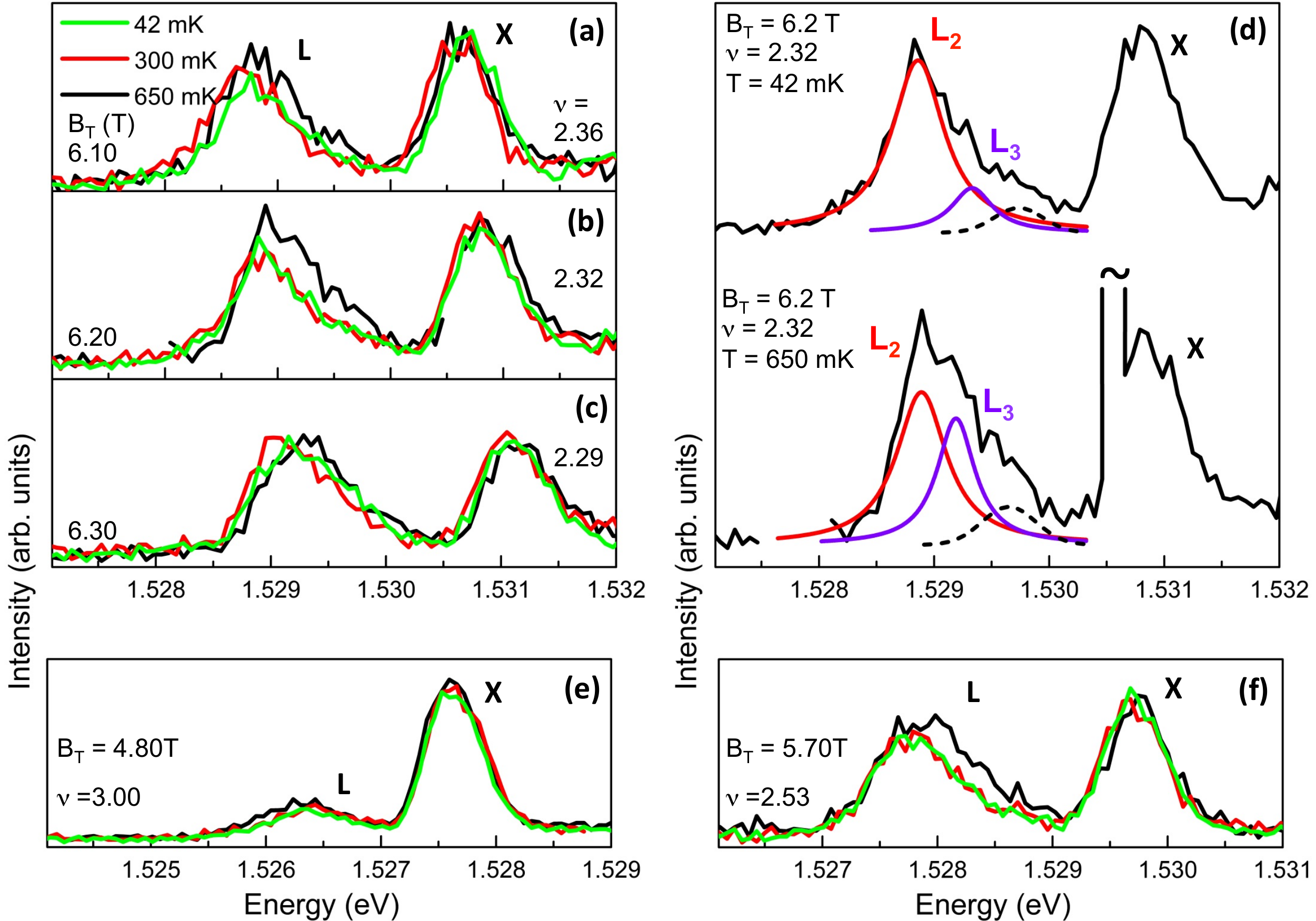}
\caption{(a-c) Temperature dependence of optical emission from sample B close to $\nu=7/3$. (d) Temperature dependence of the emission at $\nu=2.32$ (from (b)) with fits for L peaks shown. (e) Temperature dependent optical emission spectra for $\nu=3$ and (f) for $\nu=2.53$.}
\label{figS4}
\end{figure}

Figure~\ref{figS4}(a-d) illustrates the temperature dependence of the optical emission spectra from sample B at three magnetic fields for $\nu$ close to $7/3$ (a-c) and focuses on $\nu=2.32$ (d). The temperature dependence appears to be negligible for $T<$ 300 mK. The lack of temperature dependence below $T=300$~mK is characteristic of optical emission spectra throughout the filling factor range $2<\nu<3$, as exemplified by Fig.~\ref{figS4} (e,f), which show the spectra at $\nu=3$ and $\nu=2.53$ respectively. While there is clear temperature dependence at some filling factors for $T<$ 650 mK (Fig. \ref{figS4}(b,d,f)), it is not universal, and for large ranges of $\nu$, no temperature dependence is observed for $T<$ 650 mK (i.e. Fig. \ref{figS4}(e)). The fits presented in Fig.~\ref{figS4}(d) suggest competition between L$_2$ and L$_3$ as a function of temperature. The temperature dependence observed is weaker than that of FQHE and RIQHE \cite{Kumar2010} and is more similar to the weak temperature dependence of anisotropic transport \cite{Xia2011}.

\bibliographystyle{apsrev4-1}
\bibliography{Ref1bib}

\end{document}